%% file: main.tex
\documentclass[journal, 10pt]{IEEEtran}
\usepackage{soul,color}
\usepackage[nolist,nohyperlinks]{acronym}
\usepackage{xcolor}
\colorlet{correction1}{black}
\colorlet{correction2}{black}
\usepackage{tikz}
\usetikzlibrary{patterns}
\usetikzlibrary{decorations.markings}
\usepackage{float}
\usepackage[caption=false,font=footnotesize]{subfig}
\usepackage{graphicx}

\usepackage{cite}

\ifCLASSINFOpdf
\else
\fi
\usepackage{amsmath}
\usepackage[cmintegrals]{newtxmath}

\usepackage{algpseudocode}

\usepackage[colorinlistoftodos]{todonotes}
\usepackage{lipsum}
\begin{document}
%
\title{\Large Improving In-Network Computing in IoT Through Degeneracy}
%
%
%
\author{Merim~Dzaferagic, Neal~McBride, Ryan~Thomas, Irene~Macaluso, and~Nicola~Marchetti
\thanks{This publication has emanated from research supported in part by a research
  grant from Science Foundation Ireland (SFI) and is co-funded under the
  European Regional Development Fund under Grant Number 13/RC/2077 and in part
  by work supported by the Air Force Office of Scientific Research under award
  number FA9550-17-1-0066.}
\thanks{Merim~Dzaferagic, Neal~McBride, Irene~Macaluso and Nicola~Marchetti are with CONNECT, Trinity College Dublin, Ireland. (email: dzaferam@tcd.ie, mcbridne@tcd.ie, macalusi@tcd.ie, nicola.marchetti@tcd.ie)}
\thanks{Ryan~Thomas is with AFOSR, Air Force Office of Scientific Research, USA. (email: ryan.thomas@us.af.mil)}}
\maketitle

\begin{abstract}

We present a novel way of considering in-network computing (INC), using ideas from statistical physics. We define degeneracy for INC as the multiplicity of possible  options available within the network to perform the same function with a given macroscopic property (e.g. delay). We present an efficient algorithm to determine all these alternatives. Our results show that by exploiting  the set of possible degenerate alternatives, we can significantly improve the successful computation rate of a symmetric function, while still being able to satisfy requirements such as delay or energy consumption.
\end{abstract}

\begin{IEEEkeywords}
In-network computing, distributed computing, degeneracy, redundancy, IoT
\end{IEEEkeywords}

%

\section{Introduction}\label{sec:introduction}
\input{\sectionsPath/introduction.tex}

\section{Framework}\label{sec:framework}
\input{\sectionsPath/framework.tex}

\section{Analysis}\label{sec:analysis}
\input{\sectionsPath/analysis.tex}

\section{Conclusion}\label{sec:conclusion}
\input{\sectionsPath/conclusion.tex}

\input{\sectionsPath/acronyms.tex}

\ifCLASSOPTIONcaptionsoff
  \newpage
\fi



%


\bibliographystyle{IEEEtran}  
\bibliography{main}

%








\end{document}

%% file: sections/introduction.tex

\ac{IoT} networks perform actions or make decisions powered by access to raw or
processed data gathered by spatially distributed ``things''. Distributed
in-network computation (INC) approaches, i.e. the processing of raw data within
the \ac{IoT} network, are becoming increasingly popular in that they can
potentially achieve higher energy efficiency, lower computational delay, and
higher robustness compared to the traditional approach, which involves
transmitting raw data to the sink and then performing the computation.


As \ac{IoT} networks become increasingly large, their intercommunication becomes
complex enough as to resemble a thermodynamic system of many interacting
objects. This situation leads naturally to a physics-inspired study of their
behavior using statistical mechanics and a concept known as degeneracy.
Degeneracy arises from structurally different configurations of a system
(microstates) having functionally similar/identical macroscopic properties
(macrostates). A concise definition of degeneracy comes from a modern review of
the use of the term:
``degeneracy describes the ability of different structures to be conditionally
interchangeable in their contribution to system functions''
\cite{mason2015degeneracy}. In the context of INC, we view these different
structures as functional topologies (FT) \cite{Dzaferagic2016}: subsets of the
network which enable the computation of a distributed function. In this paper,
we consider degeneracy as the multiplicity of available FTs that enable the
distributed INC with a given macroscopic observable property i.e. delay.

\textcolor{correction1}{The intended application of an \ac{IoT} network usually
  defines the type of functions being considered. This in turn naturally defines
  the observables (macrostates) of interest to analyze these functions. The
  distributed computing literature
  \cite{7shah2013network,3kannan2013multi,5liu2013distributed,8vyavahare2016optimal,9yang2017graph}
  usually assumes an error free scenario, i.e. no node failures, and mainly
  focuses on the optimization of the computational rate and the communication
  cost. On the other hand, we are interested in a more realistic network setup,
  in which node failures do exist, and we study the impact of degenerate
  computation and communication paths on the success rate of the computation.}

\textcolor{correction2}{Authors like
  \cite{7shah2013network,3kannan2013multi,5liu2013distributed,8vyavahare2016optimal,9yang2017graph,di2018network,Giridhar2006}
  address communication aspects of distributed in-network computation. They
  focus on the optimization of the communication/computation parameters, like
  computational complexity, energy efficiency, computational throughput and
  computational delay. In contrast, we focus on the degeneracy of the network by
  finding the multiple feasible computational graphs that satisfy a given
  requirement (e.g. delay), rather than a single optimal computational graph.
  This way, it is possible to fully harness the computational capability of the
  network and significantly increase the robustness of the computation, while
  still being able to satisfy requirements, such as delay or energy consumption.
  Even though the INC literature features analysis of functions ranging from
  distributed neural networks \cite{di2018network} to very complex computational
  frameworks like MapReduce and Dryad \cite{5liu2013distributed}, the majority
  of \ac{IoT} applications, however, rely on simple aggregate functions like
  max, min, count, sum \cite{7shah2013network}. The distributed computation of
  these functions is usually modeled with tree structures, e.g. Steiner trees
  \cite{7shah2013network,3kannan2013multi}. Closely related to the approach
  proposed in our work is the standard Steiner Tree Packing (STP)
  problem\footnote{The Steiner Tree Packing problem is to find the maximum
    number of edge-disjoint subgraphs of a given graph that connect a given set
    of required points.} \cite{2cheriyan2007packing} and its well know
  relaxation, i.e. the Fractional Steiner Tree Problem. Instead of using the
  standard STP, we rather extend the search from minimum weight Steiner trees to
  all existing Steiner trees in which the sum of the weights on the links is
  lower than a predefined value.}

The main contributions of this work are: an efficient algorithm that generates
all functional topologies satisfying a given delay requirement, for any physical
topology; a formal definition and calculation of degeneracy and the related
concept of redundancy in terms of distributed computing over \ac{IoT} networks;
a comparison of degenerate INC and a traditional multi-hop scheme that selects a
single graph for the computation showing that it is possible to significantly
increase the probability of successful computation by exploiting degeneracy.

%% file: sections/framework.tex
\textcolor{correction2}{We define a physical network $G=(V,E)$, as a simple graph that
  contains no self-edges or repeated edges, where the set of nodes $V$
  represents the physical nodes in the network (e.g. sensors in an IoT network),
  and the set of edges $E$ represents the physical links between the nodes
  that can interact directly with each other.} Mathematical functions, $f$, can
be represented as subgraphs $H$ of the overall network which we refer to as FTs.
\textcolor{correction1}{Even though the definition of FTs is much broader
  \cite{Dzaferagic2016}, in terms of INC the FTs are directed, rooted trees,
  i.e. Steiner trees, with all edges pointing towards the root, $Y$.}
This root/sink node is where the result of a particular function $f$ must reach.
The leaves of this in-tree are the inputs of the function, $X$. These inputs may
take the form of entries in a distributed database or measurements of the
environment such as temperature. The set of inputs, $x_i \in X$, are generated
in nodes $v_i$.

The individual operations of a given function, $f(X)$, are mapped to specific
nodes in the network. Each FT, $H$, is some subset of $U \subset V$ and $D
\subset E$ in which each node and edge is involved in computing and routing
$f(X)$. Multiple, unique FTs which model the same function, $f(X)$, can be
chosen from the same physical topology, $G$. Any distinct FTs that perform the
same calculation and result in the same observable macrostate of the network
function are considered to be degenerate. The overall delay associated with a
given FT is the principal macrostate of interest in this paper. We define the
delay as the maximum graph distance between the sink node, $Y$, and any other
node, $u \in H$. Alternative macrostates of potential interest are energy
efficiency, which could be approximated by the number of edges in FT $H$, and
computational throughput (number of function calculations per unit delay).

In \cite{Giridhar2005}, the authors made the distinction between two
classes of network functions, namely divisible and indivisible. Denote a set
$W = \{ 1, \ldots, |W|\}$ and a subset $S = \{i_1, \ldots, i_k\} \subset W$,
where $i_1, < i_2, <\ldots, < i_k$. Let $x_S$ be a set $\{ x_{i_1}, x_{i_2},
\ldots, x_{i_k}\}$. A function, $f: x_S \rightarrow y$, is divisible if given
any partition of $S$, $\Pi(S) = \{s_1, \ldots, s_j\}$ of $S \subset W$, there
exists a function $g^{\Pi{(S)}}$ such that for any $x_S$,

\begin{align}
f(x_S) = g^{\Pi{(S)}} (f(x_{s_1}), \ldots, f(x_{s_k})).
\end{align}
Otherwise, $f$ is indivisible.
\vspace{-1em}

\subsection{Degeneracy of Functional Topologies} \label{subsec:weak_degeneracy}
Degeneracy can be considered in terms of these two classes of network functions.
We define strong degeneracy to be the multiplicity of FTs of an indivisible
network function. We define weak degeneracy to be the multiplicity of FTs of
divisible functions. The aggregate functions mentioned in
Sec~\ref{sec:introduction} are all examples of divisible functions. Due to the
previously discussed importance of these aggregate functions for distributed
computing, we focus on the aspects of weak degeneracy in this paper.

For a physical topology, $G$, a network function, $f$, and an input \& output
set of nodes, $X$ \& $Y$, we can define \textit{weak degeneracy} of the physical
topology, $G$, with respect to an observable (e.g., delay, energy efficiency) as
the multiplicity of FTs,
\begin{align}
g_W(G, X, Y, f, d) = |\{H(G, X, Y, f, d)\}|,
\end{align}
with a given delay, $d$. For completeness, \textit{strong degeneracy} is defined
as the multiplicity of FTs on top of a physical topology $G$ with respect to an
observable, with input nodes $X$, performing an indivisible function in which
the sub-operations must be performed in an exact ordering, $g_S(G, X, Y, f, d)$.

Since $f(X)$ is divisible, we can perform the calculation using any partition
$s$ of $X$ and combine the result of each subset $f(x_s)$ in any order. Weak
degeneracy \textcolor{correction1}{of a full mesh physical topology} is
therefore dependent on the number of ways of partitioning a set, known as the
Bell number,
\begin{align}
B_{n+1} = \sum_{k=0}^{n} \binom{n}{k} B_n,
\end{align}
for a set of $n + 1$ elements and $B_0 = B_1 = 1$.
\textcolor{correction2}{It should be noted that the Bell number is an upper
  bound for the weak degeneracy. The weak degeneracy of a generic physical
  topology has to further account for the number of mappings of these partitions
  to a given FT. This is the number of ways we can route each partition such
  that the suboperation $f(x_{s_i})$ on partition $s_i$ occurs when the elements
  of partition $s_i$ intersect for the first time.}

Our degeneracy analysis allows us to both identify the set of subgraphs which
perform a function and to sort these with regard to observables like delay,
energy efficiency or computational throughput.
This insight may be used in future to improve computational throughput,
computational resilience, deploy superior network configurations, or
to analyze the degeneracy potential of a routing protocol, i.e., how many of the
feasible FTs can a routing protocol discover.
\vspace{-1em}
\subsection{Redundancy of Functional Topologies}
Degeneracy arises from structurally different FTs having functionally identical
properties.
We can also define redundancy in FTs, which is related to degeneracy, with two
redundant FTs not sharing any common nodes or edges apart from their common $X$
and $Y$. The redundancy between two FTs, $r(H_a, H_b)$, is defined as
\begin{align}
r(H_a, H_b) = \begin{cases} 1 \text{, if } U_a \cap U_b = X \cup Y,\\
							 0 \text{, otherwise.}
	\end{cases}
\end{align}
We define the average redundancy,
\begin{align}
R(G, Y, f) = \frac{1}{| \{ X \} |}\sum_{\{X\}} \frac{1}{N_X} \underset{i \neq j}{\sum_{\{H_i,H_j\}}} r(H_i, H_j),
\end{align}
the mean over the number of FTs which are redundant to any other given FT. The
sum is over all $N_X$ possible FTs and the set of all possible input sets,
$\{X\}$. The total redundancy gives a measure of the number of ways to choose
alternate FTs which bypass nodes in case of node/link failure.

\subsection{Find all weakly degenerate functional
  topologies} \label{subsec:alg_weak_degeneracy}
In the context of INC, an FT describes the interactions between network nodes in
the course of computing a distributed function \cite{Dzaferagic2016}.
\textcolor{correction1}{For the INC of divisible functions FTs are Steiner
  trees, i.e. they include only the nodes and edges from the physical topology
  that are involved in the computation and/or routing of the function.}

To generate an FT, consider a partition $S$ of the input set $X$. For each
subset $s \in S$, the function $f$ acts on
the elements in $s$ at their first meeting point node in $H$. For a subset $s =
\{x_1, x_2\}$, we construct paths, $p_1 = x_1, v_1, \ldots, v_i, \ldots, Y$ and
$p_2 = x_2, u_1, \ldots u_k, \ldots, Y$. The sub-operation $f(x_1, x_2)$ occurs
where these paths meet for the first time, $v_l = u_k$. Paths $p_1$ and $p_2$
must contain at least one crossing point since they have a common root at $Y$.
Having met at $u_k = v_l$, the result of sub-operation $f(x_1, x_2)$ can be
routed down either sub-path to $Y$. If there exist further subsets $s$ in the
partition, these are treated in the same manner. The order in which we combine
paths between any subsets, $s$, and $Y$ does not matter since $f$ is divisible.

\begin{figure}[t]
	\centering
	\includegraphics[scale=0.25]{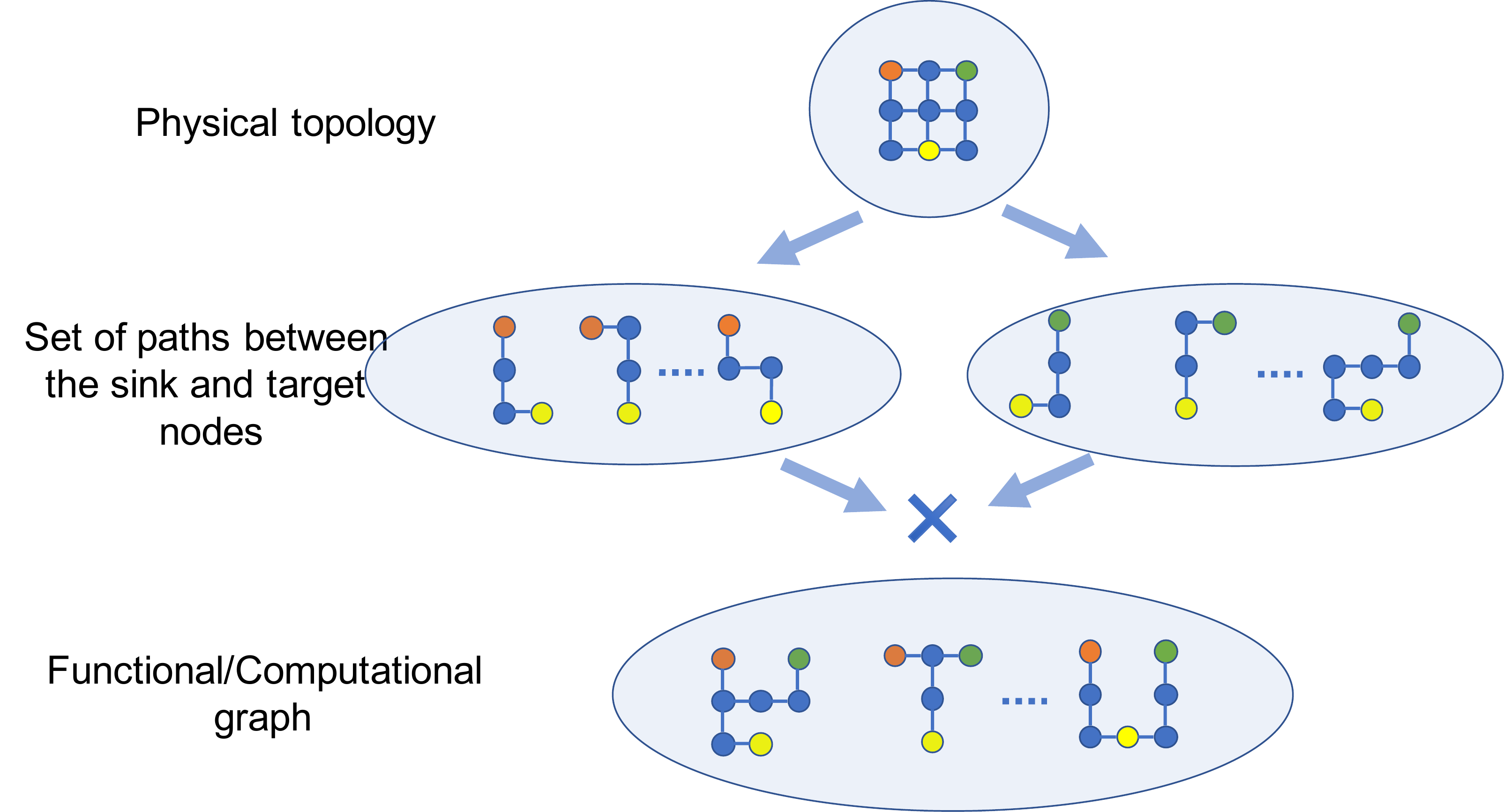}
	\caption{Process to identify all FTs from a generic physical topology.}
	\label{fig:functional_top_from_physical_alg}
  \vspace{-1.3em}
\end{figure}

\begin{figure}[h]
	\begin{algorithmic}[1]
    	\Procedure {findFTs}{$G$, $X$, $Y$, $f$, $d_{\mathrm{max}}$}
		\State Find all directed, simple paths, $\{p_{x_i}\}$, from all source nodes, $X$, to the target node, $Y$ of length $d_{\mathrm{max}}$ or less. \Comment{e.g. Breadth-First Search}
        \State Take the union of edges and nodes of all possible combinations, $C$, of paths from different input nodes $p_{x_i} \forall x_i$.
        \If {A given combination, $c\in C$, is not a tree (contains one or more cycles).}
        	\State Find all unique spanning trees of $c$.
            \For {All unique spanning trees}
            	\If {Spanning trees has leaf nodes, $L \nsubseteq X$. }
                \State Remove leaf nodes, $L$, since they are not an input.
            	\EndIf
            \EndFor

        \EndIf
        \State Remove any duplicate FTs.
        \EndProcedure
	\end{algorithmic}
	\caption{Our algorithm to generate all FTs of a graph $G$, for a symmetric function $f$,
    with input nodes $X$, sink node, $Y$, and a maximum delay cutoff $d_{\mathrm{max}}$.}
	\label{Alg:ft}
\end{figure}

If we consider a specific path each from $x_1$ and $x_2$ to $Y$ and take the
union of the nodes and edges in each path, we generate a subgraph of $G$ rooted
at $Y$ and with leaves $X$. If the union of the paths join and later diverge,
one or more undirected cycles have been formed. Each independent path around the
cycle results in a different FT. These FTs can be identified by finding the
unique spanning trees of the cycle-containing subgraphs and removing the
resulting leaves which are not inputs, $X$, since they don't perform a
computational or routing role.

We assume that $G$ is connected and there exist paths from each input node to
the sink. To aid in simulation, we restrict the maximum path length (delay),
$d_{\mathrm{max}}$, to the maximum graph distance between the sink and any other
node, known as the eccentricity. Figure
\ref{fig:functional_top_from_physical_alg} depicts this process of extracting
the FTs from a physical topology and Fig.~\ref{Alg:ft} outlines our algorithm.
\textcolor{correction1}{Our algorithm extends the standard Steiner Tree Packing problem and
its relaxation, i.e. the Fractional Steiner Tree Problem (which are known to be
NP-hard \cite{2cheriyan2007packing}), by extending the search from minimum weight Steiner trees to all
existing Steiner trees in which the sum of the weights on the links is lower
than a predefined value. To the best of our
knowledge, this problem has not been addressed in the literature so far.}
\vspace{-1em}

%% file: sections/analysis.tex

We now present the numerical analysis of the degeneracy and redundancy of two
different physical topologies: an $11\times11$ square lattice and a randomly
placed sensor network with the same number of nodes. In the lattice case, we
place the sink node, $Y$, in the centre of the lattice and we restrict the
maximum delay to $10$ hops, equal to the eccentricity of the sink node. In the
case of a random topology, we consider an area of $1\times 1$ km, place the sink
node at the centre of the area, and randomly distribute the remaining $120$
nodes. The edges are chosen to connect each node to its four closest
neighbors\footnote{It is worth noting that this results in a non-uniform
  distribution of the edges.}. The simulation involves calculating all FTs for
randomly sampled input node pairs, $X$, which are chosen to be within three hops
from each other, since in-network computation is typically used for data
collected by nearby sensors. To calculate the average number of degenerate FTs
with a given delay, this process is repeated $500$ times. The standard errors on
the means are estimated using bootstrap resampling of all simulations, and are
in-fact smaller than the plot markers.

\begin{figure}[t]
	\centering
	\includegraphics[trim={30 8 40 42},clip, scale=0.47]{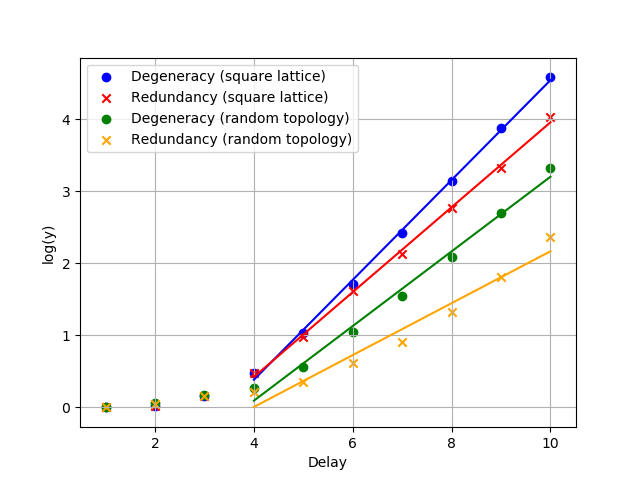}
	\caption{Log plot (base 10) of cumulative degeneracy and cumulative
    redundancy. For the square lattice, the slopes of the linear fits for the
    cumulative degeneracy and cumulative redundancy are: 0.69 and 0.59,
    respectively. For the random topology, the slopes of the linear fits for the
    cumulative degeneracy and cumulative redundancy are 0.51 and 0.36,
    respectively.}
	\label{fig:fit_deg_red_size_11_11_d_10}
  \vspace{-1.3em}
\end{figure}

Shown in Fig.~\ref{fig:fit_deg_red_size_11_11_d_10} are the cumulative
degeneracy of FTs and cumulative average redundancy for both topologies. For the
lattice case, the cumulative number of degenerate FTs (blue markers) is seen to
increase exponentially with the increasing delay limit. This same exponential
behaviour has also been seen in independent experiments on different lattice
sizes. It is a result of the branching process associated with FTs of increasing
size gaining access to more and more nodes and their respective edges. The
cumulative redundancy of FTs (red markers) also increases exponentially with
delay limit.  Although the cumulative degeneracy and redundancy are lower in the
case of a random topology than those observed in a lattice, they still increase
exponentially as the delay limit increases. 
The increase in cumulative redundancy demonstrates that the ability for physical
topologies to perform parallel computing using independent FTs is higher for
less restrictive delay limits.

\begin{figure}[t]
	\centering
	\includegraphics[trim={15 8 40 40},clip,scale=0.47]{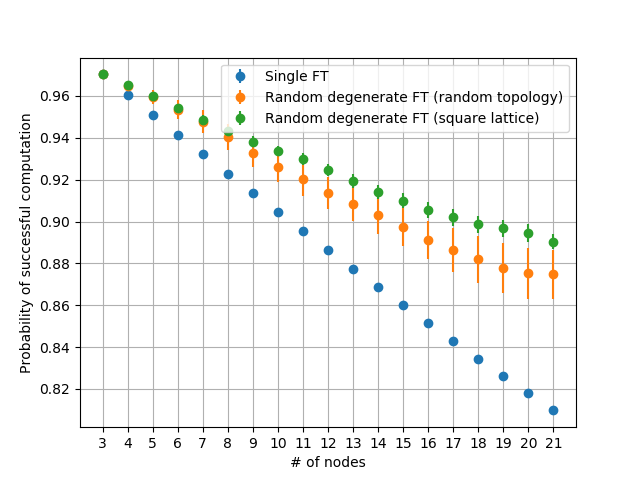}
	\caption{Probability of successful computation vs number of nodes involved in the computation. The blue dots refer to a single FT; the green and orange dots refer to a randomly selected FT that can perform the same function for the square lattice and random
    topology respectively. The delay limit for all calculations is set to be
    equal to $10$.}
\label{fig:probability_of_successful_computation}
\vspace{-1.3em}
\end{figure}

Using the numerical estimates of degeneracy, we now examine the computational
robustness of INC. 
Let us first define the probability of computational success, $\alpha_i$, at
node $i$ as the number of successful computations
as a fraction of attempts. The probability of a node failing during the computation of a function is then found to be,
\begin{equation}
  \label{eq:probability_of_node_failure}
  P_{f_i} = P_{c_i} \cdot (1-\alpha_i),
\end{equation}
where, $P_{c_i}$, is the probability of a node occurring in any feasible FT. A computation may fail at a node for a number of reasons such as excessive load or during a sleep cycle. The probability of a successful computation is given as
\begin{equation}
  \label{eq:probablity_of_succ_comp}
  P_{S_c} = \prod_{i \in U} (1 - P_{f_i}),
\end{equation}
where $U$ are the nodes in the FT that performs the computation. Figure
\ref{fig:probability_of_successful_computation} compares the probability of
successful computation using a single
FT, e.g. selected as the optimal computational graph in terms of delay or energy
consumption, and a random selection of one of the multiple FTs that can perform
the same computation while satisfying a maximum delay requirement ($10$ for the
results in figure). In the case of a single optimal FT, $P_{c_i} = 1$ for all
nodes, since all nodes in the optimal graph must be used for INC.
\textcolor{correction2}{Considering the degeneracy of FTs from
  Fig.~\ref{fig:fit_deg_red_size_11_11_d_10}, each computation can
  potentially be performed by using any of the degenerate FTs. Therefore, we
  estimate $P_{c_i}$ for the randomly selected FT, that is going to perform the
  computation, as the frequency of occurrences of the node $i$ in all the
  degenerate FTs. This $P_{c_i}$ is then used to compute the corresponding
  $P_{S_c}$ according to (\ref{eq:probability_of_node_failure}) and (\ref{eq:probablity_of_succ_comp}).}


Results are then averaged over all the $500$ input
pairs. The results in Fig.~\ref{fig:probability_of_successful_computation} show
that it is possible to significantly increase the probability of successful
computation by exploiting the multiplicity of FTs, i.e. the degeneracy.
\vspace{-0.5em}

%% file: sections/conclusion.tex
We presented a novel way of considering INC using ideas from statistical
physics. In particular, we defined degeneracy and the related concept of
redundancy for distributed computation in networks, and we introduced an
algorithm to efficiently compute all degenerate and redundant functional
topologies.
Our results show that the cumulative degeneracy and redundancy of functional
topologies increase exponentially as the accepted delay limit for the
computation is increased. The successful computation rate of a symmetric
function is shown to be significantly higher using the set of possible
degenerate functional topologies as opposed to exclusively using the optimal
functional topology.
This means that we can considerably improve the robustness of the computation,
while still being able to satisfy requirements, such as delay or energy
consumption. Future work will focus on exploiting degenerate and redundant FTs
to perform parallel computing. The case of redundant, i.e. independent FTs, is
of particular interest in that bottlenecks can be avoided with  no coordination
between the nodes of different FTs.
\vspace{-0.5em}

%% file: sections/acronyms.tex
\begin{acronym}
  \acro{IoT}{Internet of Things}
  \acro{FFT}{Fast Fourier Transform}
  \acro{SVD}{Singular-Value Decomposition}
\end{acronym}